\documentclass[twocolumn,showpacs,preprintnumbers,amsmath,amssymb]{revtex4}

\usepackage{graphicx}
\usepackage{dcolumn}
\usepackage{bm}
\usepackage[usenames]{color}

\begin{document}

\preprint{APS/123-QED}

\title{Spin-polarized Dirac-cone-like surface state with $d$ character at W(110)}

\author{K. Miyamoto$^{1}$, A. Kimura$^{2}$, K. Kuroda$^{2}$, T. Okuda$^{1}$, K. Shimada$^{1}$, H. Namatame$^{1}$, M. Taniguchi$^{1,2}$ and M. Donath$^{3}$}
 \affiliation{
$^{1}$Hiroshima Synchrotron Radiation Center, Hiroshima University, 2-313 Kagamiyama, Higashi-Hiroshima 739-8526, Japan
 }
 %


\affiliation{
$^{2}$ Graduate School of Science, Hiroshima University, 1-3-1 Kagamiyama, Higashi-Hiroshima 739-8526, Japan
}

\affiliation{
$^{3}$ Physikalisches Institut, Westf\"alische Wilhelms-Universit\"at M\"unster, Wilhelm-Klemm-Strasse 10, 48149 M\"unster, Germany
}

\date{\today}

\begin{abstract}
The surface of W(110) exhibits a Dirac-cone-like state with $d$ character within a spin-orbit-induced symmetry gap. As a function of wave vector parallel to the surface, it shows nearly massless energy dispersion and a pronounced spin polarization, which is antisymmetric with respect to the Brillouin zone center. In addition, the observed constant energy contours are strongly anisotropic for all energies. This discovery opens new pathways to the study of surface spin-density waves arising from a strong Fermi surface nesting as well as $d$-electron-based topological properties. 


\end{abstract}

\maketitle
Spin manipulation without external magnetic fields is a key requirement to revolutionize functionalities of current electronic devices.
Spin-split energy-band structures induced by strong spin-orbit coupling in nonmagnetic materials such as Rashba systems and topological insulators have attracted great attention for their promising dissipationless spin current transport~\cite{Bychkov84,Datt90, Nita97, Miron10}. 
In particular, spin characters of surface states have been discussed in detail. 
New fascinating ideas are based on helical spin structures connected to Dirac-cone-like surface states~\cite{Hasan10}.
So far, $sp$-electron materials with a single Dirac cone have been extensively studied. Although Ir oxide has been predicted to have non-trivial $d$-derived states \cite{Shitade09}, it has not been confirmed experimentally so far. Thus a material with $d$-derived topologically non-trivial states remains yet to be explored.
Possible strong correlation effects among $d$ electrons in topological insulators could be important scientific targets.
Angle-resolved photoemission spectroscopy (ARPES) with spin resolution has been proved to be the most powerful tool to characterize the spin-split energy bands of the low-index surfaces of metals and topological insulators \cite{Hoesch04, Hirahara08, Kadono08, Dil09, Hasan10}.

Tungsten is well known as the prototype material for spin-orbit coupling effects appearing, e.g., in spin- polarized low-energy-electron diffraction (LEED), which has been utilized for analyzing the electron spin polarization for decades \cite{Kirschner79}.  
The substantial spin-orbit coupling of the tungsten substrate promotes a chiral magnetic order in the adsorbed Mn lattice through Dzyaloshinsky-Moriya interaction as demonstrated by spin-polarized scanning tunneling microscopy \cite{Bode07}. 
There have been extensive ARPES studies on surface states of clean and hydrogen-covered W(110) ~\cite{Gaylord87, Gaylord88, Gaylord89, Rotenberg98, Hochstrasser02, Shikin08,Rotenberg08}.
In the latter case,  Rashba-type spin-split bands were observed~\cite{Hochstrasser02}.

  
In this letter, we show the first experimental evidence of nearly massless and strongly spin-polarized surface states in a spin-orbit induced symmetry gap of W(110). Our spin- and angle-resolved ARPES study reveals that the spin polarization is antisymmetric with respect to the $\bar{\Gamma}$ point.
~The constant-energy cuts of this Dirac-cone-like state are found to be strongly flattened compared with what has been observed in the $sp$-electron-based surface Dirac cone at the (111) surface plane of some Bi-based chalcogenides~\cite{Xia09, Chen09, Kuroda10}.
These findings can be a platform for future studies on strongly correlated $d$-electron-based topological insulators.


   A clean surface of W(110) was obtained by repeated cycles of heating in an oxygen atmosphere ($2 \times 10^{-6}$ to $2\times 10^{-7}$Pa) at 
   1500~K and a subsequent flash to 2300~K. During the flash, the pressure stayed below $5 \times 10^{-7}$ Pa. 
   This cleaning procedure was effective to remove contaminants such as carbon and oxygen from the surface as confirmed by Auger electron spectroscopy as well as a very sharp (1$\times$1) LEED pattern. 
 The ARPES experiment was performed by using linearly polarized undulator radiation on the beamline (BL-1) of the Hiroshima Synchrotron Radiation Center (HiSOR). 
The electric field vectors can be switched between parallel ($p$-polarization) and perpendicular ($s$-polarization) to the plane spanned by the surface normal and photoelectron propagation vectors as schematically shown in the left part of Fig. 1 (a).
   The overall experimental energy and angular resolutions were set to 15~meV and 0.1$^{\circ}$, respectively. 
   The spin-resolved ARPES experiment was performed with a high-flux He-discharge lamp and with a home-built Mott-type spin polarization detector attached to a commercial hemispherical electron-energy analyzer \cite{Iori06}. 
   Total energy and angular resolutions were set to 100 meV and 2$^{\circ}$, respectively. 
   The geometry of the spin-resolved ARPES experiment is described in the right part of Fig. 1 (a). 
   The angle of light incidence was $50^{\circ}$ relative to the lens axis of the electron analyzer. 
   The spin polarimeter can measure the spin component normal to the plane spanned by two vectors directed along light propagation and photoelectron emission. The photoelectron emission angle $\theta$ is measured from the surface normal.  The sign of $\theta$ is defined as positive (negative) when the surface normal is moved toward (away from) the light propagation vector. 
All measurements have been performed at a sample temperature of 100 K.
\begin{figure}
\includegraphics{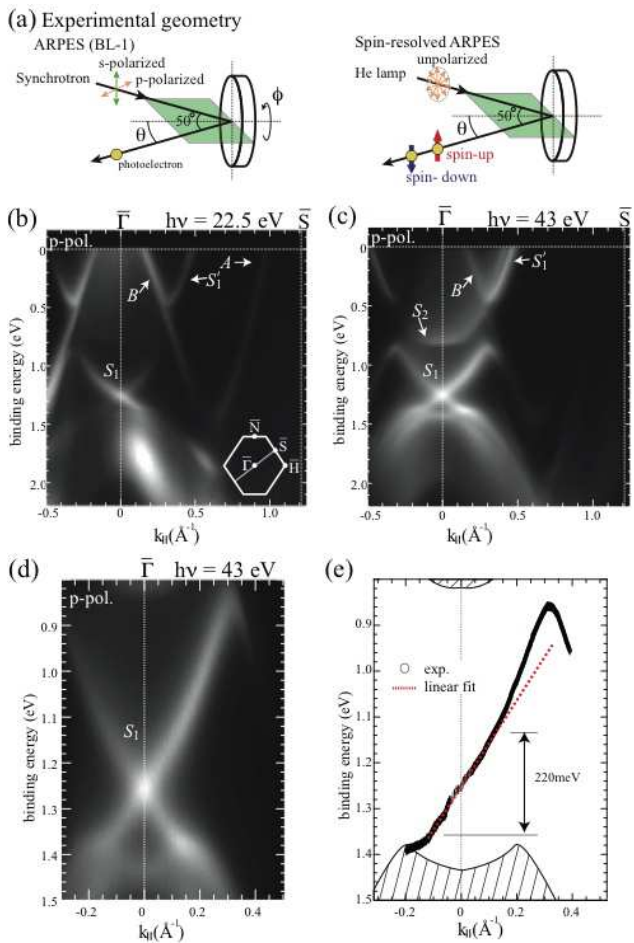}
\caption{\label{fig:epsart}(color online) (a) Schematic illustration of experimental geometries for ARPES using synchrotron radiation ($left$) and spin-resolved ARPES using a He discharge lamp ($right$). (b),(c) Energy band dispersion curves of W(110) along $\overline{\Gamma \rm S}$ measured with $p$-polarized synchrotron radiation light of $h\nu$ = 22.5~eV and 43~eV. (d) Magnified ARPES energy dispersion curve with enhanced momentum resolution for the crossing point of  $S_{1}$. (e) Intensity maxima plot for one branch of $S_{1}$ obtained from momentum and energy distribution curves in (d).  The hatched area in Fig. 1(e) shows the bulk continuum states outside the symmetry gap.} 
\end{figure}



 The energy dispersion curves taken at $h\nu$ = 22.5 eV along the $\overline{\rm \Gamma S}$ line (Fig. 1(b)) show  four characteristic surface states ($A$, $B$, $S_{1}$, $S'_1$). 
 As interpreted in a former ARPES study, the surface states $A$ 
 and $B$ originate from the surface dangling bonds~\cite{Rotenberg98}.
 We proved their surface character by exposing the surface to small amounts of adsorbates, which caused the states to disappear. 
 The presence of these states 
 thus confirms the surface cleanliness 
 in the present work. 
We note here that the surface band $S_{2}$ appears (partly near $S'_{1}$)
 for $h\nu$ = 43 eV, while it is absent for $h\nu$ = 22.5 eV.

Figure 1(d) shows the ARPES energy dispersion curve measured with enhanced momentum resolution for $S_{1}$ and Fig. 1(e) highlights its intensity maxima plot of one branch of $S_1$ obtained from momentum and energy distribution curves in the limited momentum space. The hatched area in Fig. 1(e) marks the bulk continuum states outside the symmetry gap. The surface state $S_{1}$ shows almost massless (linear) energy dispersion around the $\bar{\Gamma}$ point (-0.1~$\rm \AA^{-1}$~$\leq$~$k_{\parallel}$~$\leq$0.1~$\rm \AA^{-1}$) and a Dirac-point-like band crossing at 1.25~eV. It starts to deviate from the linear behaivor from 1.1~eV above the crossing point (upper branch) and it touches the bulk continuum states at 1.4~eV for the lower branch. Note that the linear band dispersion is persistent in the wide energy region of $\sim$ 220~meV (1.14-1.36~eV). This situation is even larger than the most studied topological insulator Bi$_{2}$Se$_{3}$ ($\sim$ 150 meV)  \cite{Kuroda101}. 

It is found that the binding energy ($E_B$) of $S_{1}$ does not change as a function of incident photon energy, which confirms its two-dimensional nature and is consistent with the previous normal-emission result \cite{Gaylord87}.  
 In addition, the previous work suggests that this band $S_{1}$  exists in the symmetry gap of the projected bulk band induced by spin-orbit interaction. 
 The present work is focused on this surface state $S_{1}$, which exhibits a Dirac-cone-like band dispersion. 
 
\begin{figure*}
\includegraphics{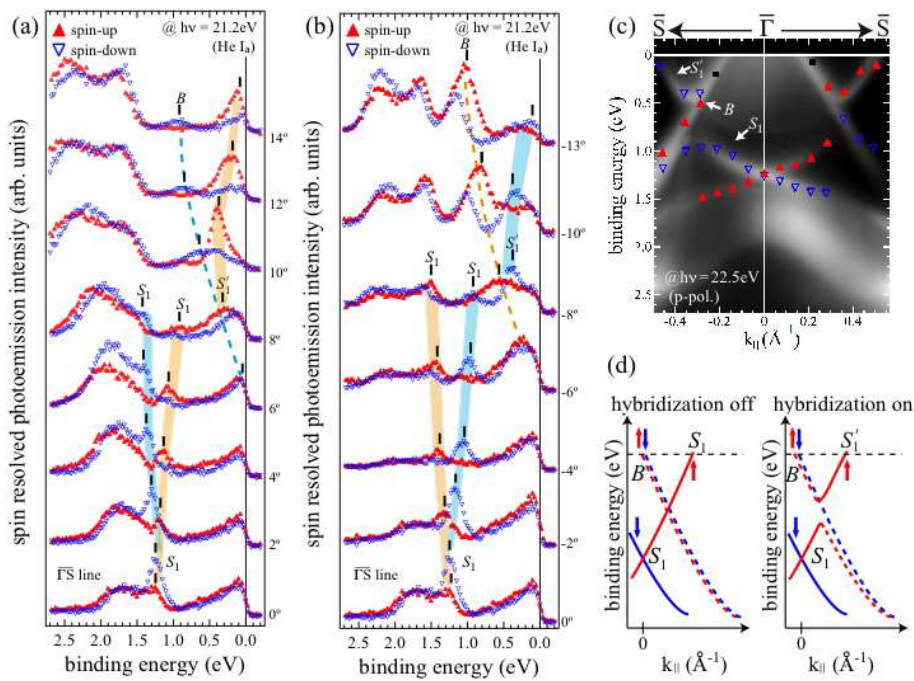}
\caption{\label{fig:epsart}(color online) (a), (b) Spin-resolved energy distribution curves (EDCs) of W(110) for 
positive and negative emission angles along the $\overline{\Gamma \rm S}$ line excited with unpolarized He-I$_{\alpha}$ light ($h\nu$ = 21.2 eV). Spin-up and spin-down intensities are marked with triangles pointing up (red) and down (blue). (c) ARPES results obtained by $p$-polarized synchrotron radiation light ($h\nu$ = 22.5 eV). The contour plot is superimposed by triangles pointing up and down indicating the spin character of the corresponding spectral features, as derived from spin-resolved spectra in Figs. 2(a) and 2(b). 
Here, the non spin-polarized peaks are denoted by squares (black). 
(d) Schematic $E$ vs $k_{\parallel}$ dispersion of $S_{1}$ and $B$ states without and with spin-selective hybridization.}
\end{figure*}

To unravel the spin characters of these nearly massless energy dispersions, we have performed spin-ARPES measurements. 
Figures 2(a) and 2(b) show the spin-resolved energy distribution curves (EDCs) in the region of $E_{B}$ = 0 to 2.7 eV for 0$^{\circ}\le\theta\le$+14$^{\circ}$ and 0$^{\circ}\ge\theta\ge$-13$^{\circ}$.
 Here, the spin-up and spin-down spectra are plotted with triangles pointing up and down, respectively.
 In the EDC for normal emission, three features are observed at binding energies ($E_{B}$) of 1.8~eV, 1.25~eV and just below the Fermi level ($E_{\rm F}$) in both spin channels.
 The broad feature centered around 1.8 eV shifts to higher $E_{B}$ as $|\theta|$ increases, while the feature at $E_{\rm F}$  does not shift for $|\theta|$ $\leq$ 4$^{\circ}$. 
Since these features are broad, almost spin degenerate, and change as a function of $h\nu$ (see ARPES results), they can be assigned to bulk continuum states.
For the prominent peak $S_{1}$, the spin-up and spin-down features appear at the same $E_{B}$ (= 1.25 eV), yet with different intensities, at $\theta$ = 0$^{\circ}$.  
More importantly, the spin-up peak shifts to lower $E_{B}$ with increasing $\theta$ from 0$^{\circ}$ to 8$^{\circ}$, while the spin-down peak shifts to higher $E_{B}$ (Fig. 2(a)). For negative angles, we observed the same $E$(k$_{\parallel}$) dispersion  but with opposite spin polarization (Fig. 2(b)). 
This result clearly shows that the $S_{1}$ band is spin split and the spin orientation is antisymmetric with respect to the $\bar{\Gamma}$ point, which is reminiscent of the surface Dirac cones of 3D topological insulators.

Let us now take a closer look at the $S_{1}$ band.
In going from 0$^{\circ}$ to +8$^{\circ}$ (-8$^{\circ}$), the spin-down (spin-up) intensity gets gradually smaller, is merged into the bulk state and its spin polarization is weakened.
The spin-up (spin-down) component 
shows small intensity until $\theta$ reaches +8$^{\circ}$ (-8$^{\circ}$) and the intensity is enhanced 
at +10$^{\circ}$ (-10$^{\circ}$) (see peak $S'_1$ at $E_{B}$ = 0.4~eV) and approaches $E_{\rm F}$ as shown in Fig. 2 (a) (Fig. 2(b)). 
In Fig. 2(c), we have assigned spin characters to the dominant spectral features in the contour plot of Fig. 1(b) by up- and down-triangles. The assignments are based on the dominant spin polarization of the features in Figs. 2(a) and 2(b).

Now we remark on the photoemission intensity of  $S_{1}$.
In the spectra obtained with unpolarized light, the spin-down intensities are larger than the spin-up intensities for $|\theta|$ $<$ 8$^{\circ}$, in particular for $\theta = 0^{\circ}$. This spectral intensity modulation is consistent with that in the spin-integrated measurements.
To find out  the reason for this intensity asymmetry, we performed ARPES measurements with $p$- and $s$-polarized light ($h\nu$ = 22.5 eV).
The result for $s$-polarized light (not shown here) exhibits comparable intensities for both branches,
 whereas one branch (predominantly spin-down) has much larger intensity for excitation with $p$-polarized light (Fig. 2(c)). 
 It is known that the use of different light polarizations ($s$ or $p$) and/or different light incidence angles for unpolarized light result in different photoemission intensities. These experimental parameters enter the matrix elements for the optical transitions. The square of the matrix elements, in turn, determines the experimentally observed intensity \cite{Oepen86,Lindroos02}. We thus conclude that the asymmetric intensities with respect to $\bar{\Gamma}$ in Figs. 2(a) and 2(b) originate from the matrix-element effects as a consequence of the experimental geometry. 

\begin{figure*}
\includegraphics{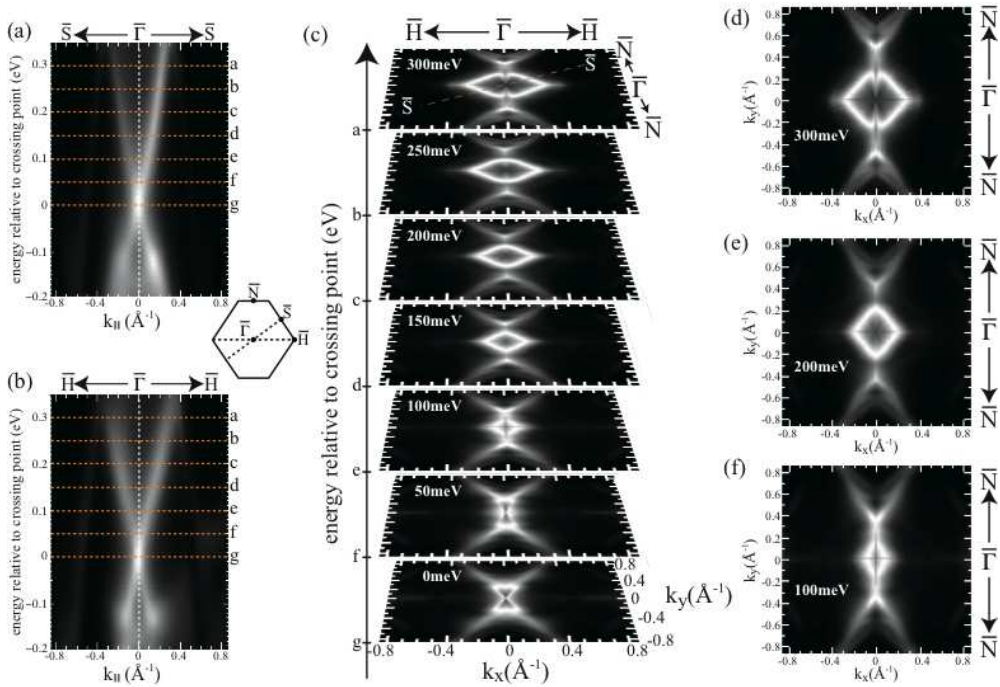}
\caption{\label{fig:epsart} ARPES results obtained with $p$-polarized synchrotron radiation light ($h\nu$=43~eV).
(a), (b) Energy dispersion curves along $\overline{\Gamma \rm S}$ and $\overline{\Gamma \rm H}$.
(c) Constant energy contours at several energies from 0  to 300 meV above the crossing point. 
(d)-(f) Selected constant energy contours at 300 meV, 200 meV and 100 meV. }
\end{figure*}

We notice 
from Figs. 1(c) and 2(a-c) that there is a spin-dependent hybridization between the $S_{1}$ and the $B$ band near $k_{\parallel}$ = 0.3~\AA$^{-1}$ ($E_{B}$ = 0.85~eV) as schematically shown in Fig. 2(d).
At low $E_B$ far from the hybridization point, the state $B$ does not show spin polarization.
 $B$ moves to higher $E_{B}$ with increasing $|\theta|$ and, depending on the sign of $\theta$, acquires negative or positive spin polarization for $|\theta|$ $>$ 8$^{\circ}$, where the $S_{1}$ and $B$ bands cross each other.
Since the upper cone of S$_{1}$ shows only one spin component for each $k_{\parallel}$ direction, 
only one spin component of $B$ can hybridize. Consequently, the other spin component of $B$ does not hybridize with $S_{1}$ (see Fig. 2 (d)). 
As a result of the avoided crossing and the hybridization-gap formation in one spin component, two peaks appear at $E_{B}$  = 0.4  ($S'_1$) and 1.0~eV ($S_1$) in the spin-up and spin-down channel for $\theta$ = 8$^{\circ}$ and -8$^{\circ}$, respectively (see the spin resolved EDCs at $\theta$ = $\pm$8$^{\circ}$ in Figs. 2(a) and 2(b)).  
A careful analysis of the intensity behavior of $S_1$ with increasing $|\theta|$ shows decreasing intensity upon approaching the hybridization gap
, followed by a transfer of intensity to the $S^{'}_{1}$ branch.
A similar case of spin-selective hybridization was reported for quantum well states in Bi/Ag(111) \cite{He10}.

The present interpretation is supported by previous reports on the H-covered W(110) surface.
Once the dangling bonds are terminated by H atoms,  both surface states $A$ and $B$ disappear.
The former report shows that  $S_{1}$ becomes connected to the surface band $S'_1$ in the absence of hybridization with state $B$~\cite{Rotenberg08}.
A further photoemission study on H-W(110) employed spin resolution~\cite{Hochstrasser02}. It was shown that the surface states on the hydrogen-covered surface were indeed gspin split and spin-polarized by the spin-orbit interaction in association with the loss of inversion symmetry near a surfaceh. The spin polarization was measured for two different bands corresponding to $S_{1}$ and $S_{2}$ in our present notation, yet modified by hydrogen adsorption as described above. Due to the limited momentum and energy resolution ($\sim$300~meV) at that time, the spin polarization at the lower branch was hardly resolved. In addition, a linear dispersion behavior was not identified in the former work. In the present work,  the distinct spin polarization at both upper and lower branches is clearly revealed (see Fig. 2(a)-(c)), as well as a linear dispersion behavior with a crossing point at $\bar{\Gamma}$. Both observations provide clear evidence of Dirac-cone-like states at the clean W(110) surface.

Two-dimensional slices in $\bf{k}$ space (-0.87 $\rm \AA^{-1}$ $\leq$ $ k_{x}$, $ k_{y}$ $\leq$ 0.87 $\rm \AA^{-1}$) at several constant energies 
starting from the crossing point at $E_{B}$=1.25~eV are shown in Fig. 3. 
The whole set of images has been acquired within one hour in order to avoid a possible energy shift by charge transfer from  
adsorbates as observed in other systems \cite{Hsieh09}. 
In fact, we have observed an energy shift of up to 100 meV for $S_{1}$, when the sample 
was measured in ultra-high vacuum for more than six hours. 
A diamond-shaped constant energy contour is observed at 200 meV above the crossing point.
Upon approaching the crossing point, the size of the contour becomes smaller. 
At higher energy from the crossing point, we find a distorted hexagon in the constant energy contour, see Fig. 3(d) at 300~meV. 
This shape reflects the two-fold symmetry of the W(110) surface.
 There are pairs of parallel lines in the constant energy contours along the $\rm{\overline{\Gamma S}}$ direction. The shape of the constant-energy contour is highly suggestive of spin-density-wave formation derived from strong nesting
as discussed for Bi$_{2}$Te$_{3}$ and Bi$_{2}$Se$_{3}$~\cite{Fu09, Kuroda101}.
Moreover,  we find quite anisotropic group velocities of $v_{g} = (1.4 \pm 0.1) \times 10^{5}$ and (1.6 $\pm$ 0.1) $\times 10^{5}$ m/s around the crossing point ($-0.1~\rm \AA^{-1} \leq  k_{\parallel} \leq  0.1~\rm \AA^{-1}$) along $\rm{\overline{\Gamma S}}$ and $\rm{\overline{\Gamma H}}$, respectively. Both of them are almost half compared to those in the $sp$-like surface states of Bi chalcogenides \cite{Hasan10, Hsieh09, Kuroda10}.


 
 In summary, a nearly massless energy dispersion curve formed by $d$ electrons at the W(110) surface has been observed by high-resolution ARPES.
The two-dimensional constant energy contours clearly show strongly anisotropic features whose size and shape change as a function of energy.
This anisotropic Dirac cone on the W(110) surface is found to be strongly spin polarized with anti-symmetric spin polarization with respect to the surface Brillouin zone center by spin-resolved measurements.
 With increasing $k_{\parallel}$, this spin-polarized massless surface band shows spin-selective hybridization with a non-polarized surface resonance, i.e., a hybridization gap opens for one spin component only.
The discovered anisotropic Dirac cone on the W(110) surface paves the way to the study of peculiar  topologies in $\bf{k}$ space with interesting spin textures. The present system provides information beyond normal $sp$-type Rashba systems and topological insulators because of the $d$-character of the involved electronic states.
	
We thank J. Braun, J. Henk, P. Kr\"uger, and J. Min\'ar for stimulating discussions and K. Kanomaru, R. Nishimura, H. Iwasawa, and H. Hayashi for experimental support. M.D. gratefully acknowledges the hospitality of the Hiroshima Synchrotron Radiation Center, and support by the Japan Society for the Promotion of Science (Invitation Program for Advanced Research Institutions in Japan). The measurements were performed with the approval of the Proposal Assessing Committee of HSRC (Proposal Nos. 10-A-27, No. 10-B-14).

\bibliography{apssamp}

\begin{thebibliography}{40}

\bibitem{Bychkov84} Y.A. Bychkov and E. I. Rashba, JETP Lett. {\bf 39}, 78 (1984).

\bibitem{Datt90} S. Datta, and B. Das, Appl. Phys. Lett. {\bf 56}, 665 (1990).

\bibitem{Nita97} J. Nitta, T. Akazaki, H. Takayanagi, and T. Enoki, Phys. Rev. Lett.  {\bf 78}, 1335 (1997).

\bibitem{Miron10} I. M. Miron, G. Gaudin, S. Auffret, B. Rodmacq, A. Schuhl, S. Pizzini, J. Vogel, and P. Gambardella, Nature Mater. {\bf 9}, 230 (2010).

\bibitem{Hasan10}  M. Z. Hasan, and C. L. Kane, Rev. Mod. Phys. {\bf 82}, 3045 (2010).

\bibitem{Shitade09} A. Shitade, H. Katsura, J. Kune$\check{\rm s}$, X.-L. Qi, S.-C. Zhang, and N. Nagaosa, Phys. Rev. Lett {\bf 102}, 256403 (2009).


\bibitem{Hoesch04} M. Hoesch, M. Muntwiler, V. N. Petrov, M. Hengsberger, L. Patthey, M. Shi, M. Falub, T. Greber, and J. Osterwalder, Phys. Rev. B {\bf 69}, 241401(R) (2004).

\bibitem{Hirahara08} T. Hirahara, K. Miyamoto, A. Kimura, Y. Niinuma, G. Bihlmayer, E. V. Chulkov, T. Nagao, I. Matsuda, S. Qiao, K. Shimada, H. Namatame, M. Taniguchi, and S Hasegawa, New J. Phys. {\bf 10}, 083038 (2008).

\bibitem{Kadono08} T. Kadono, K. Miyamoto, R. Nishimura, K. Kanomaru, S. Qiao, K. Shimada, H. Namatame, A. Kimura, and M. Taniguchi, Appl. Phys. Lett. {\bf 93}, 252107 (2008).

\bibitem{Dil09} J. H. Dil, J. Phys. Condens. Matter {\bf 21}, 403001 (2009).

\bibitem{Kirschner79}  J. Kirshner, and R. Feder, Phys. Rev. Lett {\bf 42}, 1008 (1979)./J. Kirschner, {\it Polarized Electrons at Surfaces} (Springer-Verlag, Berlin, 1985).

\bibitem{Bode07} M. Bode, M. Heide, K. von Bergmann, P. Ferriani, S. Heinze, G. Bihlmayer, A. Kubetzka, O. Pietzsch, S. Bl$\ddot{\rm u}$gel, and R. Wiesendanger, Nature {\bf 447}, 190 (2007).

\bibitem{Gaylord87} R. H. Gaylord and S. D. Kevan, , Phys. Rev. B {\bf 36}, 9337 (1987).

\bibitem{Gaylord88}  R. H. Gaylord, and S. D. Kevan, Phys. Rev. B {\bf 37}, 8491 (1988).

\bibitem{Gaylord89} R. H.~Gaylord, K. H.~Jeong, and S. D.~Kevan, Phys. Rev. Lett {\bf 62}, 2036 (1989).

\bibitem{Rotenberg98} E.~Rotenberg and S. D.~Kevan, Phys. Rev. Lett.  {\bf 80}, 2905 (1998).

\bibitem{Hochstrasser02}M.~Hochstrasser, J. G.~Tobin, E.~Rotenberg, and S. D.~Kevan, Phys. Rev. Lett. {\bf 89}, 216802 (2002).

\bibitem{Shikin08} A. M. Shikin, A. Varykhalov, G.V. Prudnikova, D. Usachov, V. K. Adamchuk, Y. Yamada, J. D. Riley, and O. Rader, Phys. Rev. Lett. {\bf 100}, 057601 (2008).

\bibitem{Rotenberg08}E.~Rotenberg, O.~Krupin, and S. D.~Kevan, New J. Phys. {\bf 10}, 023003 (2008).

\bibitem{Xia09} Y. Xia, D. Qian, D. Hsieh, L. Wray, A. Pal, H. Lin, A. Bansil, D. Grauer, Y. S. Hor, R. J. Cava, and M. Z. Hasan, Nature Phys. {\bf 5}, 398 (2009).

\bibitem{Chen09} Y. L. Chen, J. G. Analytis, J.-H. Chu, Z. K. Liu, S.-K. Mo, X. L. Qi, H. J. Zhang, D. H. Lu, X. Dai, Z. Fang, S. C. Zhang, I. R. Fisher, Z. Hussain, and Z.-X. Shen, Science {\bf 325}, 178 (2009).

\bibitem{Kuroda10}K. Kuroda, M. Ye, A. Kimura, S. V. Eremeev, E. E. Krasovskii, E. V. Chulkov, Y. Ueda, K. Miyamoto, T. Okuda, K. Shimada, H. Namatame, and M. Taniguchi, Phys. Rev. Lett. {\bf 105}, 146801 (2010). 

\bibitem{Iori06} K. Iori, K. Miyamoto, H. Narita, K. Sakamoto, A. Kimura, S. Qiao, K. Shimada, H. Namatame, and M. Taniguchi, Rev. Sci. Instrum. {\bf 77}, 013101 (2006).






\bibitem{Oepen86} H. P. Oepen, K. H$\ddot{\rm u}$nlich, and J. Kirschner,  Phys. Rev. Lett. {\bf 56}, 496 (1986).

\bibitem{Lindroos02} M. Lindroos, S. Sahrakorpi, and A. Bansil, Phys. Rev. B. {\bf 65}, 054514 (2002).

\bibitem{He10} K. He, Y. Takeichi, M. Ogawa, T. Okuda, P. Moras, D. Topwal, A. Harasawa, T. Hirahara, C. Carbone, A. Kakizaki, and I. Matsuda, Phys. Rev. Lett. {\bf 104}, 156805 (2010).


\bibitem{Hsieh09}  D. Hsieh, Y. Xia, D. Qian, L. Wray, J. H. Dil, F. Meier, J. Osterwalder, L. Patthey, J. G. Checkelsky, N. P. Ong, A. V. Fedorov, H. Lin, A. Bansil, D. Grauer, Y. S. Hor, R. J. Cava, and M. Z. Hasan, Nature {\bf 460}, 1101 (2009).

\bibitem{Kuroda101} K. Kuroda, M. Arita, K. Miyamoto, M. Ye, J. Jiang, A. Kimura, E. E. Krasovskii, E. V. Chulkov, H. Iwasawa, T. Okuda, K. Shimada, Y. Ueda, H. Namatame, and M. Taniguchi, Phys. Rev Lett {\bf 105}, 076802 (2010).



\bibitem{Fu09} L. Fu, Phys. Rev. Lett. {\bf 103}, 266801 (2009).

%
\end{thebibliography}

\end{document}